\documentclass[iop,apjl]{emulateapj}

\begin{document}

\newcommand{\grb}{GRB~140515A}
\newcommand{\lya}{Ly$\alpha$}
\newcommand{\lyb}{Ly$\beta$}
\newcommand{\lyg}{Ly$\gamma$}
\newcommand{\taueff}{\ensuremath{\tau_{\mathrm{GP}}^{\mathrm{eff}}}}
\newcommand{\ew}{\ensuremath{EW_{\mathrm{r}}}}
\newcommand{\lnh}{\ensuremath{\log(N_{\mathrm{H I}})}}
\newcommand{\lnhcm}{\ensuremath{\log(N_{\mathrm{H I}}, \mathrm{cm}^{-2})}}
\newcommand{\zgrb}{\ensuremath{z_{\mathrm{GRB}}}}
\newcommand{\xh}{\ensuremath{\bar{x}_{\mathrm{H I}}}}

\shorttitle{GRB 140515A at $z=6.33$}
\shortauthors{Chornock et al.}

\title{GRB 140515A at $z$=6.33: Constraints on the End of Reionization
  From a Gamma-ray Burst in a Low Hydrogen Column Density Environment
} 

\author{Ryan Chornock\altaffilmark{1},
Edo Berger\altaffilmark{1},
Derek B. Fox\altaffilmark{2,3},
Wen-fai Fong\altaffilmark{1},
Tanmoy Laskar\altaffilmark{1},
and Katherine C. Roth\altaffilmark{4}
}

\altaffiltext{1}{Harvard-Smithsonian Center for Astrophysics,
                 60 Garden Street, Cambridge, MA 02138, USA; 
                 \texttt{rchornock@cfa.harvard.edu}}
\altaffiltext{2}{Department of Astronomy and Astrophysics,
  Pennsylvania State University, 525 Davey Laboratory, University
  Park, PA 16802, USA}
\altaffiltext{3}{Sabbatical address:  Max Planck Institute for
  Extraterrestrial Physics, 
  Giessenbachstrasse, 85741 Garching, Germany}
\altaffiltext{4}{Gemini Observatory, 670 North Aohoku Place, Hilo, HI
  96720, USA}

\begin{abstract}
We present the discovery and subsequent spectroscopy with Gemini-North
of the optical afterglow of the {\it Swift} gamma-ray burst (GRB)
140515A.  The spectrum exhibits a well-detected continuum at 
wavelengths longer than 8915~\AA\ with a steep decrement to zero
flux blueward of 8910~\AA\ due to \lya\ absorption at
redshift $z\approx6.33$.  Some transmission through the \lya\ forest
is present at $5.2<z<5.733$, but none is detected at higher
redshift, consistent with previous measurements from quasars and GRB
130606A. 
We model the red damping wing of \lya\ in three ways that provide
equally good fits to the data: (a) a single host galaxy absorber at
$z=6.327$ with \lnhcm=18.62$\pm$0.08; (b) pure intergalactic medium
(IGM) absorption from $z=6.0$ to $z=6.328$ with a constant neutral
hydrogen  
fraction of $\xh=0.056^{+0.011}_{-0.027}$; and (c) a hybrid 
model with a host absorber located within an ionized bubble of radius
10 comoving Mpc in an IGM with $\xh=0.12\pm0.05$ (\xh$\lesssim$0.21 at
the 2$\sigma$ level). 
Regardless of the model, the sharpness of the dropoff in transmission
is inconsistent with a substantial neutral fraction in the IGM at this
redshift. 
No narrow absorption lines from the host galaxy are detected,
indicating a host metallicity of [Z/H]~$\lesssim-0.8$.
Even if we assume that all of the hydrogen absorption is due to the
host galaxy, the column is unusually low for a GRB sightline, similar
to two out of the other three highest-redshift 
bursts with measured \lnh.  This is possible evidence that the escape
fraction of ionizing photons from normal star-forming galaxies
increases at $z\gtrsim6$.
\end{abstract}

\section{Introduction}

The reionization of the intergalactic medium (IGM) by the first stars
and quasars was likely a complex process.  Current evidence from
observations of polarization of the cosmic microwave background
implies a redshift for reionization of $z\approx10$ \citep{wmap},
while tracers at lower redshift suggest that the tail end of the
process was still ongoing at redshifts $z\approx6$ \citep{fanaraa}.    
Gamma-ray bursts (GRBs) offer one of the most promising probes of this
epoch due to their bright afterglows, which can be seen at
cosmological distances and lack some of the complications of quasar
observations (e.g., \citealt{lamb00,bl04}).

In this {\it Letter}, we present our Gemini spectrum of the afterglow
of the long-duration \grb. It reveals a  
sharp decrement in flux shortward of $\sim$8900\AA, implying a
redshift of $z$$\approx$6.32  \citep{gcn16269}, which we refine
below.  This makes \grb\ the burst with the third largest
spectroscopic 
redshift.  Only GRBs 080913 ($z$=6.733; \citealt{greiner09,patel10})
and 090423 ($z$$\approx$8.23; \citealt{tanvir09,salva}) are known to be
more distant, although GRB~090429B has a photometric
redshift of $\sim$9.4 \citep{nino_z9}.
 It is important to note 
that at the time of our Gemini observations, no optical observations
or photometric redshift estimates had been reported in the GCN
Circulars, so the high-$z$ nature of the burst was purely
serendipitous.

We describe our spectroscopic observations in
Section 2 and analyze the spectrum to set constraints on the opacity
of the IGM at $z>5.2$ in section 3.  We fit the red damping wing of
\lya\ with  three fiducial models in Section 4 to set limits on the
neutral fraction of the IGM.  We conclude in Section 5 by placing
these observations in the context of other constraints on
reionization. 

\section{Observations}

\begin{figure*}
\epsscale{1.2}
\plotone{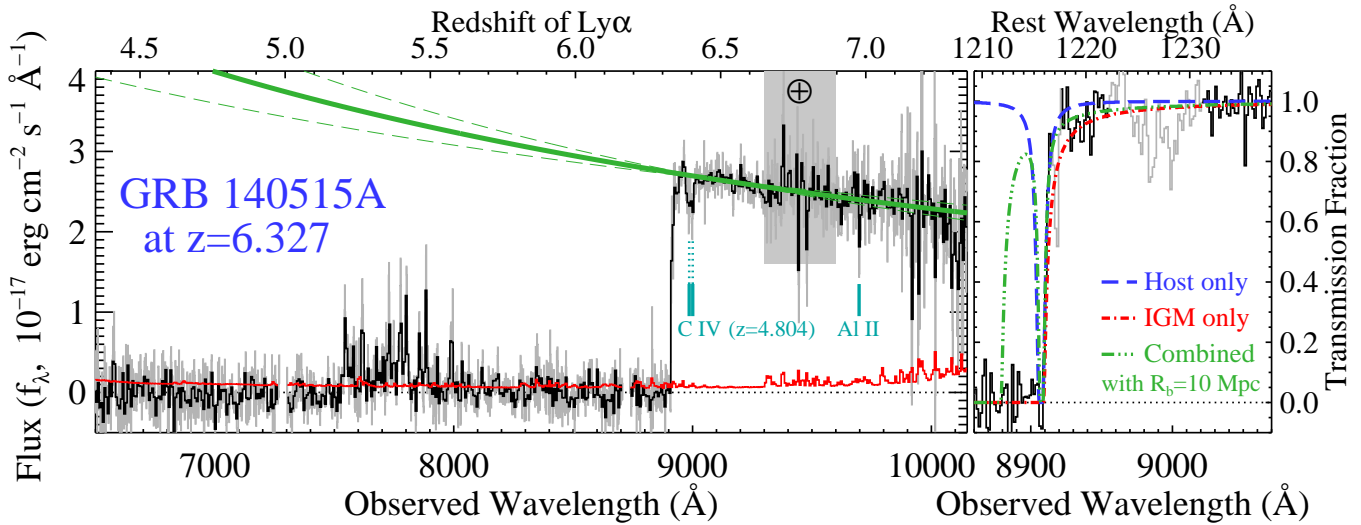}
\caption{ {\it Left:} GMOS spectrum of the optical afterglow of \grb.
  The gray spectrum is unbinned, while the black version has been
  binned
  by 5 pixels.  The red line is the 1$\sigma$ uncertainty of the
  binned spectrum.  The solid green line is a power-law fit to the
  afterglow continuum and the dashed green lines represent power laws
  with slopes of $\pm$3$\sigma$.
  The gray box with the $\earth$ symbol marks the region possibly
  containing residuals from the correction for
  telluric H$_2$O absorption.  Absorption lines from a foreground
  absorber are marked in cyan.
{\it Right:} Zoomed-in view of the normalized unbinned spectrum near
\lya. The three lines represent three fits to the red damping wing of
the \lya\ profile with \zgrb\ allowed to vary (blue: absorber at
the host redshift only; red: no host contribution, but a uniform IGM
with a constant ionization fraction at $6.0<z<\zgrb$; green: a combined
model where the GRB host resides inside an ionized bubble with a 
radius of 10 comoving Mpc).  The rest wavelength scale at the top
assumes the value derived from the host-only fit of $z$=6.327.
The spectral bins in light gray were excluded from the fit. 
See text for details.
}
\label{specfig}
\end{figure*}

 \grb\ was detected by {\it Swift} at 09:12:36 UT on 
2014 May 15 \citep{gcn16267}.  We initiated followup observations with
the Gemini Multi-Object Spectrograph (GMOS; \citealt{hook04}) on the
8~m Gemini-North telescope at 10:48 UT and discovered a faint 
optical source consistent with the X-ray afterglow position
\citep{gcn16267} in our $i$ and $z$ acquisition images
\citep{gcn16274}. We obtained a dithered pair of 900~s spectra
starting at 11:42 UT on 2015 May 15 (2.5~hr after the BAT trigger;
\citealt{gcn16267}).  During the observations, the source position
increased in airmass from 2.38 to 3.00. 
We oriented the 1$\arcsec$-wide slit at a position angle of
78$\fdg$3, which was within 5$\degr$ of the parallactic angle, so
differential slit losses should be minimal.  We used the R400 grating
with an OG515 filter to cover the wavelength range 5888--10162~\AA.
The GMOS focal plane has three abutting CCDs and we were only able to
obtain observations with a single grating angle, so our data have two
gaps at 7275--7298~\AA\ and 8709--8737~\AA.  In addition to the
high airmass, the moon was $>$99\%
illuminated, so the sky background was extremely high.
 Fortunately, the seeing was exceptional, with a
full-width at half-maximum (FWHM) of 0$\farcs$45 in the
acquisition images \citep{gcn16274} and even in our final spectrum,
the spatial FWHM of the object trace was only 0$\farcs$6. 
The spectral FWHM of night sky emission lines are $\sim$7~\AA, but due
to the excellent seeing, the object did not fill the slit and the
resolution of the object spectrum is $\sim$5~\AA.

We use IRAF\footnote{IRAF is
  distributed by the National Optical Astronomy Observatories,
    which are operated by the Association of Universities for Research
    in Astronomy, Inc., under cooperative agreement with the National
    Science Foundation.} to perform basic two-dimensional image
processing, remove cosmic rays \citep{lacos}, and shift and stack
the two-dimensional frames prior to spectral extraction.
We apply flux calibrations and correct for telluric absorption using
our own IDL procedures and archival observations of standard stars
taken with a similar setup.  The airmass of our data was
substantially higher than any suitable standard star observations that
we are able to locate in the Gemini archive, so the overall spectral
shape of the final spectrum is somewhat uncertain, although this
does not affect any of our analysis.  In 
addition, we are careful to manually scale the correction for telluric
H$_2$O absorption near 9400~\AA\ because of the large airmass
difference with the available standards.  We also correct the spectrum
for $E(B-V)=0.02$~mag of Galactic extinction \citep{eddiedoug}.

The final extracted spectrum is shown in Figure~\ref{specfig}.
The sharp decrement in flux shortward of 8910~\AA\ is evidence of
\lya\ absorption at $z\approx6.33$. We fit a power law to the
continuum flux at 9020--9870~\AA, while excluding the region
of the most uncertain telluric correction near 9400~\AA\ and find a
best fit of $f_{\lambda}$$\propto$$\lambda^{-1.56 \pm 0.13}$, which is
shown as the solid green line.  The value for the slope has additional
uncertainty due to
low-level telluric absorption over most of the available continuum.

We search the afterglow continuum for narrow absorption features from
the host galaxy interstellar medium but find no plausible matches at
wavelengths compatible with the approximate redshift of 
\lya, so we are unable to make a precise measurement of the redshift
of \grb\ (\zgrb), a point that we will return to below.  Instead, we
find unresolved absorption lines at 8985.2, 9000.0, and
9697.9~\AA\ that are consistent with the \ion{C}{4} $\lambda\lambda$
1548, 1551 doublet and \ion{Al}{2} $\lambda$1670 from an intervening
absorber at $z=4.804$.  A possible weak absorption feature at
9170.4~\AA, with an observed-frame equivalent width ($EW$) of
2.1$\pm$0.8~\AA, does not match any other expected line from the
absorber or the host galaxy\footnote{We note that a foreground galaxy
  fell in our  spectroscopic slit $\sim$3$\farcs$5 away from the GRB
  afterglow.  It 
exhibits a single spectrally-resolved emission line at 8200~\AA, which
we identify as [\ion{O}{2}] $\lambda$3727 at z=1.200.  No strong
absorption lines from that galaxy are expected in our limited
continuum wavelength interval.}. 

\section{IGM Opacity}

We first examine the transparency of the IGM to \lya\ photons over the
redshift range $5.2<z<6.3$.  The signal-to-noise ratio (S/N) of
these data (peaking in the continuum at $\sim$20 per 1.38~\AA\ pixel
near 9000~\AA) are not as high as for GRB 130606A \citep{rc13} or
for typical high-$z$ quasar observations but still place meaningful
constraints.  
  We inspect the sky-subtracted
two-dimensional spectra and find no clear transmitted continuum
over the wavelength range 8185--8905~\AA\ (corresponding to redshifts
of 5.733--6.325 relative to \lya).

\begin{figure*}
\epsscale{1.2}
\plotone{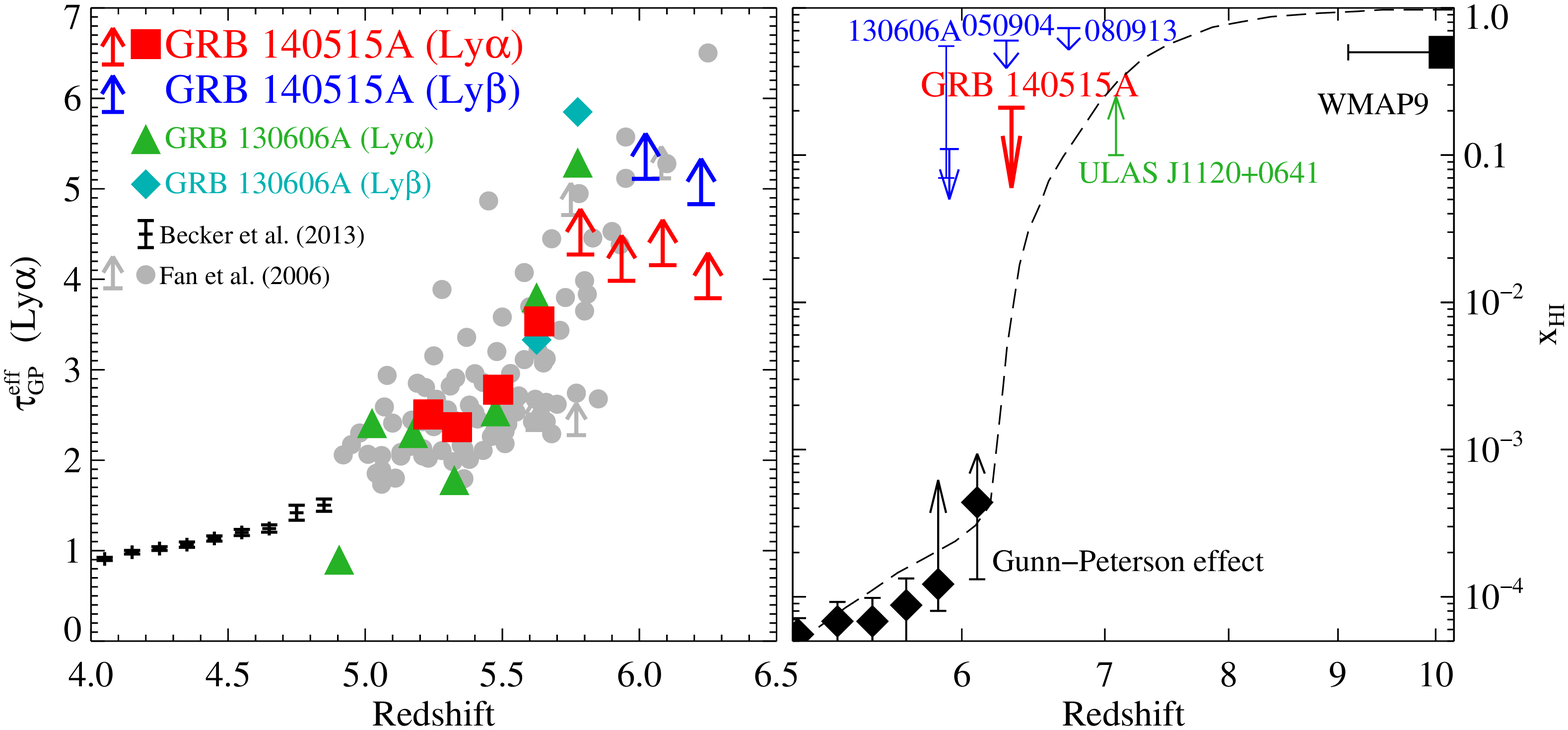}
\caption{Redshift evolution of the IGM. {\it Left:} \lya\ effective
  optical  depth,  \taueff(\lya), computed in bins of width $\Delta
  z\approx0.15$ in both \lya\ and  \lyb, 
  with arrows representing lower limits.  The triangles and diamonds
  are the measurements from the sightline to GRB 130606A
  \citep{rc13}. 
  The comparison points were measured from \lya\ 
  absorption of quasars by \citet{fan06} and \citet{becker13}.
  {\it Right:} Neutral fraction of the IGM, \xh, showing the 2$\sigma$
  limit from the hybrid IGM+host+bubble model for \grb\ in red and the
  other GRB constraints in blue
  \citep{rc13,totani06,totani14,patel10}.
  The lower limit from observations of a $z=7.085$ quasar is in
  green \citep{mortlock,bolton11}, along with 
  measurements from quasar absorption \citep{fan06} and a
  step-function reionization fit to the microwave
  background polarization \citep{wmap}.  The dashed line is a fiducial
  late reionization model from \citet{gnedin14}.
}
\label{taufig}
\end{figure*}

We follow \citet{fan06} and
define an effective optical depth to \lya, 
$\taueff = -\ln(\mathcal{T})$, where $\mathcal{T}$ is the average
transmission relative to the continuum in a bin.  
To compare to previously reported results from
$z\approx6$ quasars and GRB 130606A, we compute
\taueff\ in bins of width $\Delta z=0.15$, taking care to define our
wavelength intervals to avoid \lyb\ and the CCD gaps.  We also measure
the transmission in the \lyb\ forest and convert these to effective
\lya\ opacities using the same correction factors as \citet{fan06}. We
present the results in Figure~\ref{taufig} along with the numerical
values in Table~\ref{transtab}.  
We do not detect any transmission at $z$$>$5.71, but we are able to
set lower limits on \taueff\ of $\sim$4.
Collectively, the results from the two GRB sightlines are
generally consistent with those seen towards quasars. 

\begin{deluxetable}{lccc}
\tabletypesize{\scriptsize}
\tablecaption{\grb\ \lya\ Transmission}
\tablehead{
\colhead{Redshift Range} &
\colhead{Line} &
\colhead{Transmission} &
\colhead{\taueff (\lya)}
}
\startdata
5.20$-$5.26\tablenotemark{a} & \lya\ & 0.082 &  2.51 \\
5.26$-$5.41 & \lya\ & 0.094 &  2.37 \\
5.41$-$5.56 & \lya\ & 0.062 &  2.78 \\
5.56$-$5.71 & \lya\ & 0.029 &  3.53 \\
5.71$-$5.86 & \lya\ & $\lesssim$0.014\tablenotemark{c} & $\gtrsim$4.3 \\
5.86$-$6.01 & \lya\ & $\lesssim$0.019 & $\gtrsim$4.0 \\
6.01$-$6.16 & \lya\ & $\lesssim$0.016 & $\gtrsim$4.2 \\
6.19$-$6.31 & \lya\ & $\lesssim$0.022 & $\gtrsim$3.8 \\
5.96$-$6.08\tablenotemark{b} & \lyb\ & $\lesssim$0.103 & $\gtrsim$5.1 \\
6.15$-$6.30 & \lyb\ & $\lesssim$0.117 & $\gtrsim$4.8 
\enddata
\tablenotetext{a}{Lower redshift limit truncated to avoid
  \lyb\ absorption from host}
\tablenotetext{b}{Redshift limits set to avoid \lyg\ and a CCD chip gap}
\tablenotetext{c}{Upper limits are 3$\sigma$}
\label{transtab}
\end{deluxetable}

The large redshift bins used above effectively over-weight the
transmission to regions of lower opacity and are a crude
tracer of \lya\ opacity when the optical depth is large.  Instead,
higher-order statistics, such as the distribution of length scales of
dark gaps (e.g., \citealt{sc02}) or pixel-scale transmission
statistics \citep{mcgreer} are more sensitive probes of the structure
of reionization.
Future work with additional GRB sightlines, in combination with
numerical simulations, will enable systematic tests of the
statistics of \lya\ opacity \citep{gallerani} with a
less-biased tracer of the end of reionization than quasars, which
reside in more massive dark matter halos \citep{mes10}.

\section{\lya\ Red Damping Wing Analysis}

We now analyze the red damping wing of \lya\ to set constraints on the
neutral hydrogen column density responsible for the absorption.  We
exclude the two regions shown in gray in the right panel 
of Figure~\ref{specfig}, which include the foreground \ion{C}{4}
absorber and the residuals from an unfortunately placed strong night
sky line at 8919~\AA.  Because we do not detect any narrow lines from
the host galaxy, \zgrb\ is a free parameter in all fits.

\subsection{Pure Host Absorption}

Our first model assumes that all of the absorption is from
a single absorber, presumably the host galaxy, at redshift \zgrb.
This model is shown as the blue dashed line in Figure~\ref{specfig}.
Our best-fit parameters are \lnhcm=18.62$\pm$0.18 at
\zgrb=6.3269$\pm$0.0007.  Only 5--10\% of GRBs at $z$$>$2 have
measured host hydrogen column densities this low
\citep{chen07,fynbo09}.  

\begin{figure}
\epsscale{1.2}
\plotone{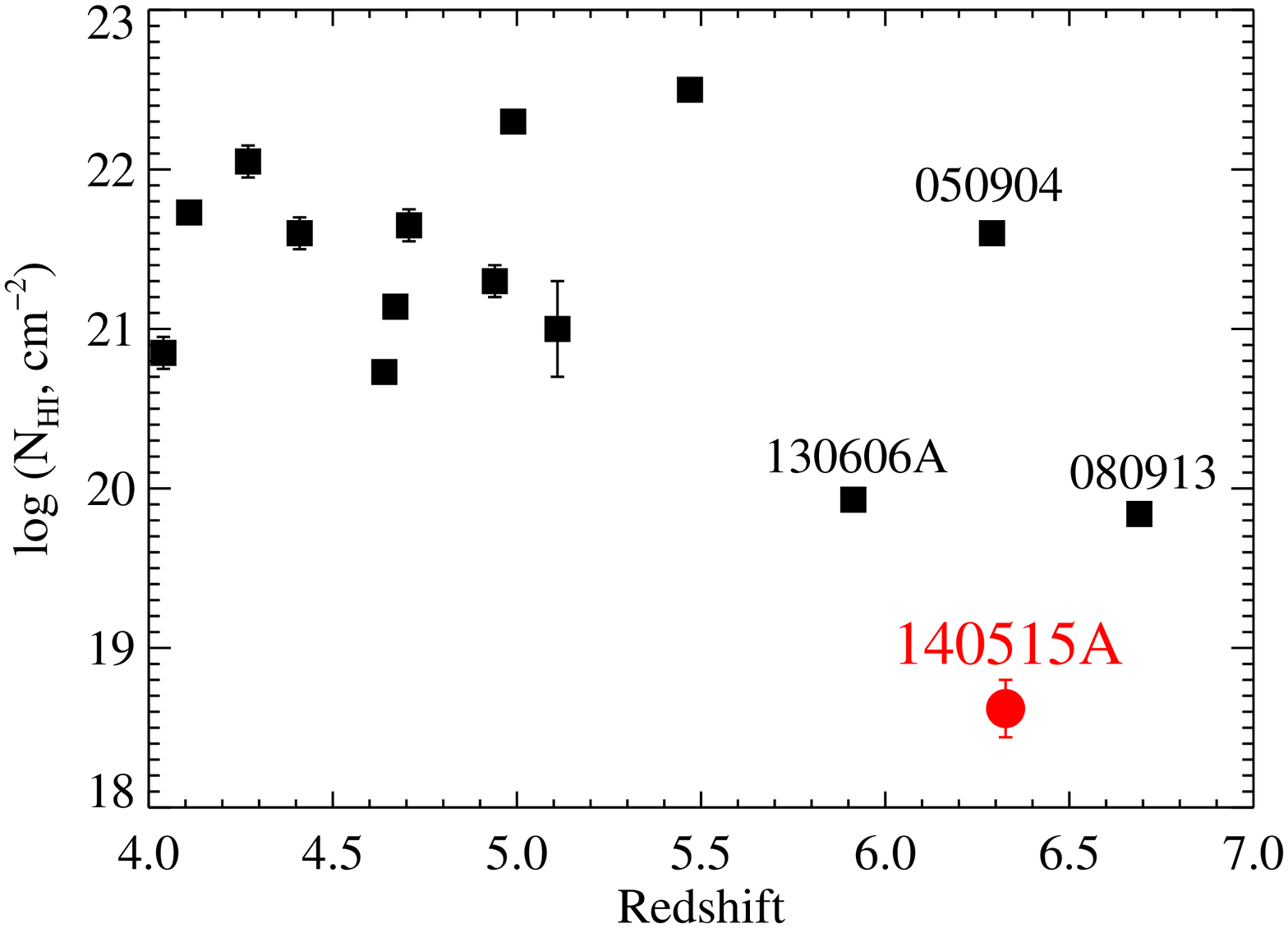}
\caption{\ion{H}{1} column density measurements for $z$$>$4 GRBs.
  Data from the 
  literature were compiled by \citet{thoene2013} and supplemented with
  a few more recent results (111008A: \citealt{sparre}; 130606A:
  \citealt{rc13,ct13,totani14}; 
  140518A: \citealt{gcn16301}).  GRBs at $z\gtrsim5.5$ appear to have
  lower  hydrogen columns than those at lower redshift.
}
\label{nhfig}
\end{figure}

\citet{chen07} argued that the distribution
of \lnh\ for a complete sample of GRBs offers a statistical sampling
of sightlines from regions of massive-star formation and can be
used to infer the escape fraction of ionizing photons in high-$z$
galaxies.  In Figure~\ref{nhfig}, we show the distribution of
available \lnh\ measurements for $z$$>$4 GRBs (mostly compiled by
\citealt{thoene2013}).  It is striking that the four highest-redshift
GRBs include the three lowest values for \lnh\ in the sample.  Lower
neutral hydrogen column densities imply a higher escape fraction.
An Anderson-Darling k-sample test gives a probability of
1.8\% that the distribution of column densities at $z>5.9$ comes from
the same distribution as that at lower redshift.  The redshift
cut was not made blindly, so more data are needed to test whether the
neutral hydrogen column distribution evolves to lower values at higher
redshift.   One
concern is selection biases, particularly that sightlines with
higher columns
will potentially be associated with more dust extinction and
thus less likely to have afterglows identified and spectra taken with
adequate S/N, but it is not clear how much of a differential effect
this could be between the $z\approx4$ and $z\approx6$
samples.

This result may be consistent with recent observations of 
a decreasing fraction of \lya-emitters among Lyman-break
galaxies at $z>6$, which have also been modeled in terms
of an increase in escape fraction (e.g., \citealt{dij14}), although an 
increase in the neutral hydrogen fraction (\xh) of the IGM (e.g.,
\citealt{schenker}) is an alternative hypothesis.  
These are complementary observations, as GRBs are superior tracers of
star formation at high redshift compared to flux-limited samples of 
galaxies \citep{tanvir12}.

Our non-detection of narrow absorption lines from the host galaxy may
be simply a consequence of the low neutral hydrogen column density.
Some  of the strongest lines we expect in the limited wavelength
interval of our continuum are \ion{Si}{2} $\lambda$1260, \ion{O}{1}
$\lambda$1302, and \ion{C}{2} $\lambda$1334.  We use the noise in the
spectrum to estimate conservative 3$\sigma$ upper limits on the
observed-frame EWs of these lines of 1~\AA, 2.5~\AA, and 2.5~\AA, 
respectively.  Assuming the optically-thin limit and the measured
value of \lnh, these correspond to upper limits on the
gas-phase abundances of [Z/H]$\lesssim$$-1.1$, $-0.6$, and $-0.8$,
respectively \citep{asplund09}. If the IGM contributes to the neutral
hydrogen column, these limits would have to be relaxed.  
Nevertheless, they are compatible with typical sub-solar abundance
estimates for high-$z$ GRB host galaxies (e.g., \citealt{thoene2013,
  rc13, ct13}) and we do not require unusually low abundances to
explain the lack of detected absorption lines from the host.

\subsection{Pure IGM Absorption}

Our second model ignores the likelihood of an absorber associated with
the host galaxy and embeds a ``naked'' GRB in a uniform medium of
constant \xh.  We use the model of \citet{jordi} to approximate the
effects of a partially neutral IGM. The model requires a lower
redshift bound for neutral hydrogen absorption, 
which we fix to $z$=6.0 because it has no effect on the results, and
an upper redshift bound that we assume is equal to \zgrb.  The
best fit is shown as the red dot-dashed line in Figure~\ref{specfig}
and has \zgrb=6.3282$^{+0.0010}_{-0.0004}$ and
\xh=0.056$^{+0.011}_{-0.027}$.  At the 3$\sigma$ level, the maximum
value allowed for \xh\ is 0.09.

\subsection{Hybrid Model}

The two models above represent possible extremes of absorption
from only the host or IGM.  If both are
present, the situation is more complex.  If we simply combine the
models, the limits will relax and favor somewhat smaller values for
\xh\ and \lnh\ as the two sources of opacity are both allowed to
contribute.  However, more properly treating the inhomogeneity
of the IGM during of the end of reionization
will considerably relax these constraints \citep{mf08,mcquinn08}.

\begin{figure}
\epsscale{2}
\plottwo{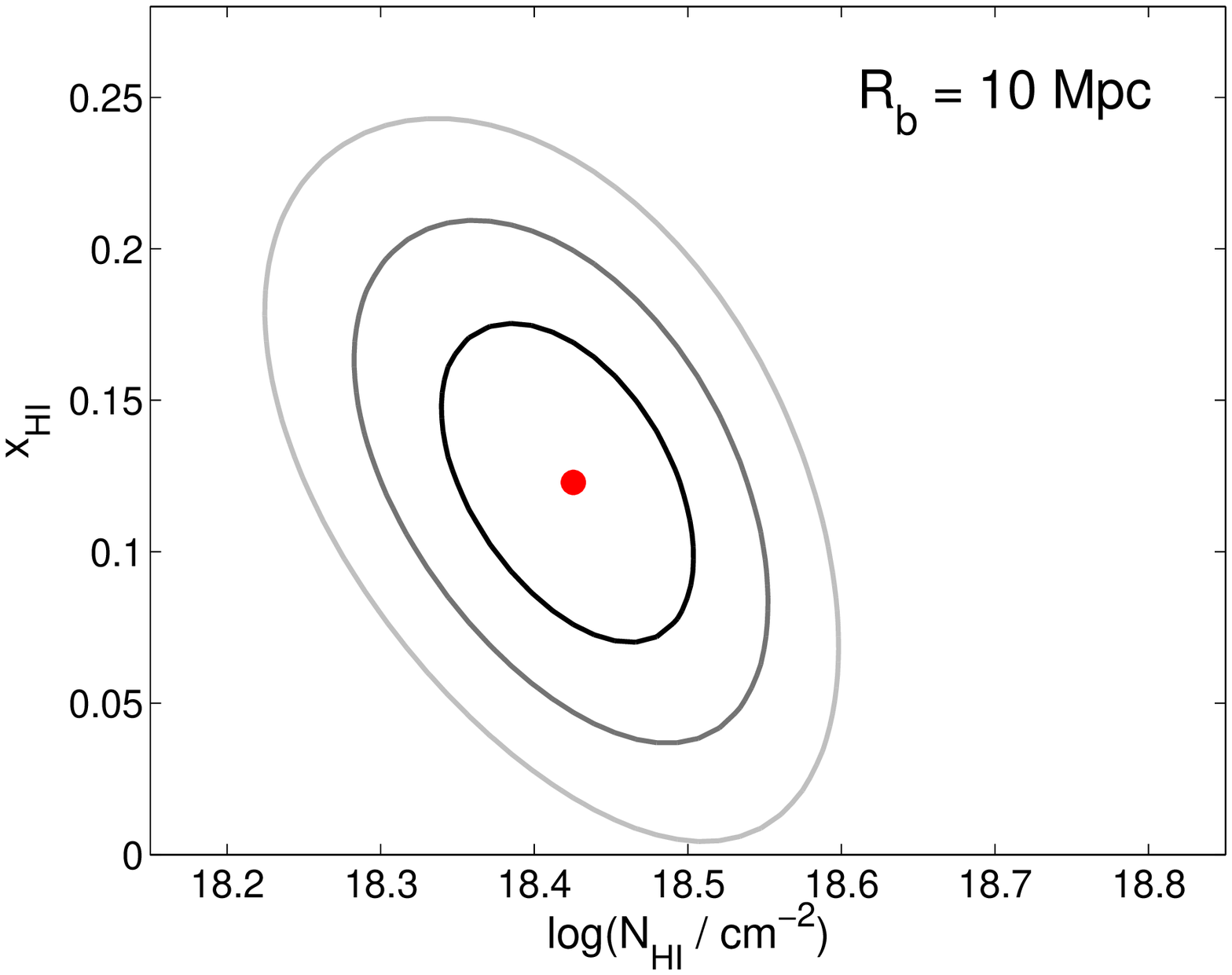}{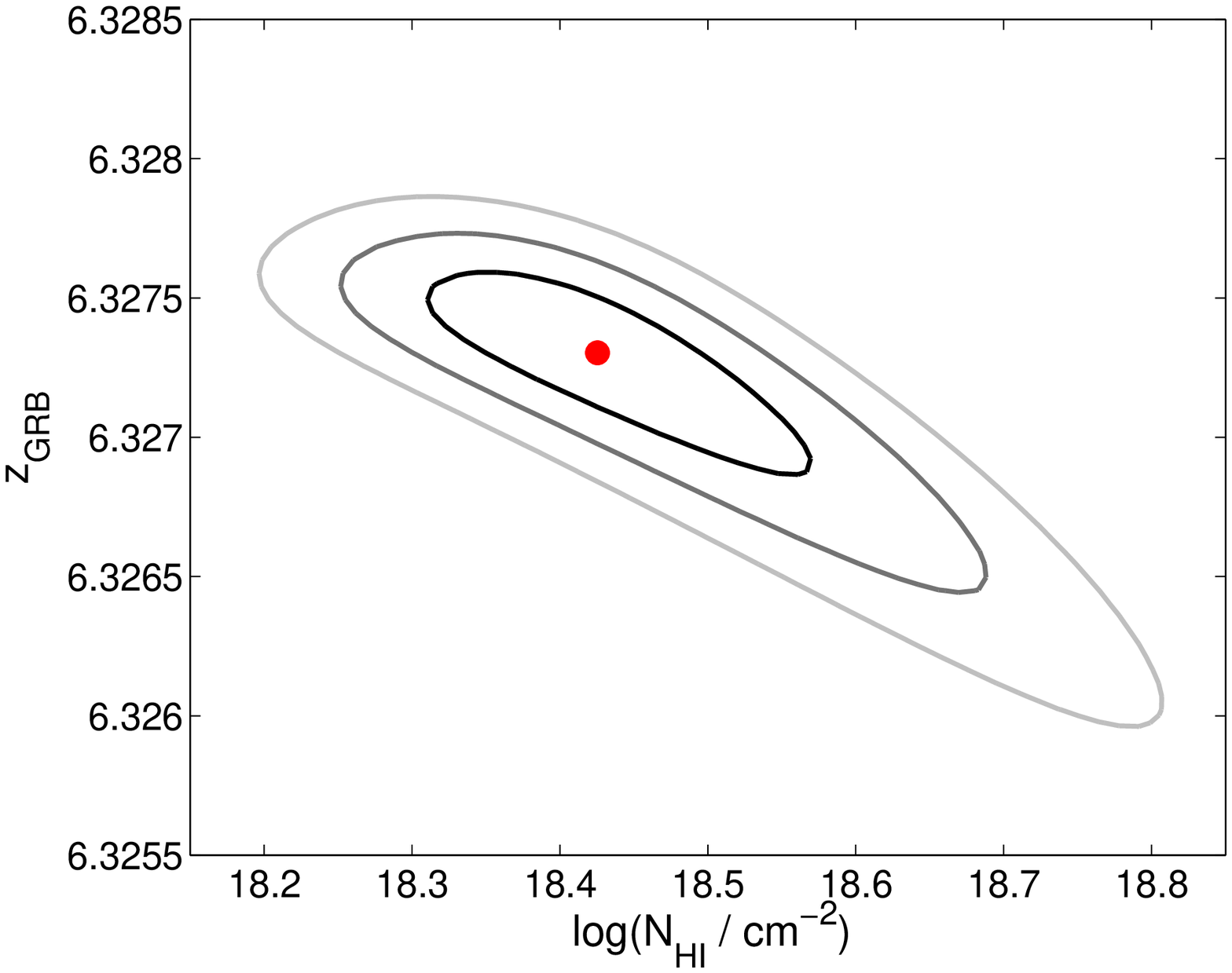}
\caption{Contours (1, 2, and 3$\sigma$) from a three-parameter fit to
  the red damping 
  wing of the \lya\ profile.  The GRB was assumed to lie at \zgrb\ in
  a host galaxy with a hydrogen column density \lnh, residing in an
  ionized bubble of radius 10 comoving Mpc in a medium with an
  otherwise uniform neutral fraction \xh, which we constrain to be
  less than 0.21 at the 2$\sigma$ level.
}
\label{igmfig}
\end{figure}

We address this issue by fitting a hybrid model with absorption from
both the 
host and the IGM.  This model has three free parameters (\lnh, \xh,
and \zgrb).  We place the GRB in an ionized bubble and fix the radius
of the bubble along our sightline to $R_{\mathrm  b}$=10 comoving
Mpc. This choice is motivated by the simulations of \citet{mcquinn08},
who found \ion{H}{2} regions of approximately this scale when
\xh\ was globally equal to 0.5.  The best-fit values are \zgrb=6.3273,
\lnh=18.43, and \xh=0.12. The marginalized contours from our fit are
shown in Figure~\ref{igmfig}.  As expected, this significantly relaxes
the constraint on \xh.  Even with the favorable
  assumption 
  about the existence of an ionized bubble, \xh$<$0.21 at the
  2$\sigma$ level.
 We do not take into account the effect of
inhomogeneity on the expected shape of the damping wing from the IGM
\citep{mf08}.  A larger bubble size would formally allow for even
higher \xh, but such bubbles are rare unless \xh\ is low
\citep{mcquinn08}.  Smaller ionized bubbles would only tighten this
constraint.   

The three fiducial models sample the range
of variation of plausible reionization scenarios.
All three models provide acceptable
fits to the data, with reduced-$\chi^2$ values near 1.0, although the
IGM-only model appears to be less adequate at the sharp edge of \lya.
The patchiness of the end of reionization will result in significant
stochasticity along different sightlines \citep{mf08,mcquinn08}
and a statistical ensemble of GRB sightlines will be necessary
to measure cosmic variance and jointly constrain the models in
more detail.

\section{Discussion and Conclusions}

We have presented an optical spectrum of the afterglow of \grb\
at $z\approx6.33$ and used it to provide constraints on the opacity and
ionization state of the IGM.

This is now the fourth high-$z$ GRB for which a damping-wing analysis
has been performed.  \citet{totani06} found that for GRB 050904 at
$z$=6.295 \citep{kawai06}, $\xh<0.6$ at the 95\% confidence level, but
that the high
host hydrogen column density inhibited tight constraints.
\citet{mcquinn08} examined the same spectrum and argued that
proper treatment of inhomogeneity will relax the constraint even
further. \citet{patel10} were unable to place meaningful constraints
on the sightline to the $z$=6.733 GRB 080913 due to the low S/N of
the data, but quoted $\xh<0.73$ at the 90\% confidence level.  The
high-$z$ GRB afterglow with the  
highest S/N observations was GRB 130606A at $z$=5.913. \citet{rc13}
set a 2$\sigma$ upper limit of \xh$<$0.11 for a model with IGM
absorption extending up to \zgrb. \citet{totani14} argued that
their spectra of the same burst disfavored a pure host-galaxy
absorption model.  If they allowed IGM absorption at redshifts up to
\zgrb, they found \xh=0.086$^{+0.012}_{-0.011}$.  Lowering the upper
redshift to $z$=5.83 allowed values for \xh\ up to $\sim$0.5.
However, their claims are based on a subtle 0.6\% deviation from
the pure host-galaxy model.

Quasars are the other major observational probe of the ionization
state of the IGM at these redshifts, but variations in intrinsic
properties and difficulties in modeling quasar near zones  
make the interpretation more controversial (e.g.,
\citealt{mh07,carilli10}).
The highest-redshift quasar currently known, ULAS J1120$+$0641
at $z$=7.085, exhibits a red damping wing
ascribed to \xh$>$0.1 \citep{mortlock}, although the
exact value depends on the unknown details of the quasar's lifetime
\citep{bolton11}. 

In our toy model with an $R_{\mathrm b}$=10 Mpc (comoving) bubble, we
set a 2$\sigma$ 
upper limit of \xh$<$0.21, with a best-fit value of \xh=0.12.
Models with smaller or nonexistent \ion{H}{2} regions will prefer even
smaller values for \xh.
The sharp decrement in flux at \lya\ is incompatible with a large
neutral fraction in the IGM at $z$$\approx$6.3 in almost any model.
We compare this value to the others described above in the righthand
panel of Figure~\ref{taufig}.  The GRB constraints on \xh\ are still
above predictions from models of the IGM tuned to reproduce the
\lya\ forest at $z\lesssim6$ (e.g., \citealt{gnedin14}).

The reasonable quality of the constraint on \xh\ from these data
despite the short integration time is a result of the brightness of
the afterglow and the low hydrogen column density of the host.  A
future sample of GRBs at even higher redshift, if it can be collected,
would serve as an excellent record of the end of reionization.

\acknowledgments
We thank the Gemini staff, particularly Lucas Fuhrman, for their
superb assistance in obtaining these observations. 
The Berger GRB group at Harvard is supported by the National Science
Foundation under Grant AST-1107973. 
Based in part on observations obtained under Program ID
GN-2014A-Q-38 (PI: Berger) at the Gemini Observatory, which is
operated by the Association of Universities for  
    Research in Astronomy, Inc., under a cooperative agreement with the
    NSF on behalf of the Gemini partnership: the National Science
    Foundation (United States), the Science and Technology Facilities
    Council (United Kingdom), the National Research Council (Canada),
    CONICYT (Chile), the Australian Research Council (Australia),
    Minist\'{e}rio da Ci\^{e}ncia, Tecnologia e Inova\c{c}\~{a}o (Brazil)
    and Ministerio de Ciencia, Tecnolog\'{i}a e Innovaci\'{o}n Productiva
    (Argentina).  

{\it Facilities:} \facility{Gemini:Gillett (GMOS-N)}

\end{document}